\documentclass[11pt]{article}
\usepackage{moriond,epsfig}
\usepackage{axodraw}

\def\Journal#1#2#3#4{{#1} {\bf #2}, #3 (#4)}

\def\NPB{{\em Nucl. Phys.} B}
\def\PLB{{\em Phys. Lett.}  B}
\def\PRL{\em Phys. Rev. Lett.}
\def\PRD{{\em Phys. Rev.} D}

\def\ra{\rightarrow}

\def\be{\begin{equation}}
\def\ee{\end{equation}}
\def\bea{\begin{eqnarray}}
\def\eea{\end{eqnarray}}

\def\lgl{\langle}
\def\rgl{\rangle}

\begin{document}
\title{$B^{0}_{s}-\overline{B^{0}_{s}}$ MIXING FROM LATTICE $QCD$}

\author{Beno\^\i t BLOSSIER }

\address{Laboratoire de Physique Th\'eorique, Universit\'e Paris XI,\\
91405 Orsay Cedex, France}

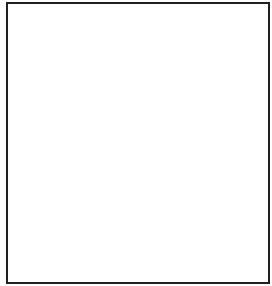
\maketitle\abstracts{
We study the $B^0_s-\overline{B^0_s}$ mixing amplitude in Standard Model by 
computing  the relevant hadronic matrix element in the 
static limit of lattice HQET with the Neuberger light quark action. In the quenched 
approximation, and after matching to the $\overline{\rm MS}$ scheme in QCD, we obtain 
$\hat{B}^{\overline{\rm MS},{\rm NLO}}_{B_s}(m_b)=0.940(16)(22)$.}

$B^0_s-\overline{B^0_s}$ mixing is highly important in testing the Standard 
Model (SM) and constrains strongly its extensions. Since it is a flavor changing 
neutral process, it occurs through loops so that the corresponding mixing amplitude 
is a sensitive
 measure of $|V_{ts}|$ and $|V_{tb}|$, as the major SM loop contribution comes from $t$ 
quark. The mixing of weak interaction eigenstates $B^0_s$ and 
$\overline{B^0_s}$ induces a mass gap $\Delta M_s$ between the mass eigenstates 
$B_{sH}$ and $B_{sL}$: experimentally, only a lower bound to $\Delta M_s$ is known currently, 
namely $\Delta M_s > 14.4 \; \rm{ps}^{-1}$ at 95 \% CL. \cite{PDG} 

Theoretically the $B^0_s-\overline{B^0_s}$ mixing is described by means of an Operator 
Product Expansion, $i.e.$ the Standard Model Lagrangian ${\cal L}_{SM}$ is reduced to an effective 
Hamiltonian ${\cal H}_{eff}^{\Delta B=2}$, plus negligible terms of ${\cal O}(1/M^2_W)$:

\be
{\cal H}^{\Delta B=2}_{eff}=\frac{G^2_F}{16\pi^2}M^2_W 
(V^*_{tb}V_{ts})^2\eta_B S_0(x_t) C(\mu_b)Q^{\Delta B=2}_{LL}(\mu_b)\quad\quad 
\mu_b\sim m_b
\ee
where $\eta_B=0.55\pm 0.01$, \quad
$S_0(x_t)=\frac{4x_t-11x^2_t+x^3_t}{4(1-x_t)^2}-\frac{3x^3_t\ln
(x_t)}{2(1-x_t)^3}$, \quad $x_t=\frac{m^2_t}{M^2_W}$  .

$C(\mu_b)$ is computed perturbatively, at NLO in $\alpha_s(\mu_b)$ in the 
$\overline{\rm MS}$ (NDR) scheme; $Q^{\Delta B=2}_{LL}$ is a 
four-fermions operator coming from the reduction of the box diagrams in 
${\cal L}_{SM}$ to a local operator in the effective theory. 
The hadronic matrix element of $Q^{\Delta B=2}_{LL}$ is conventionnally parametrized 
as
\be
\lgl\overline{B^0_s}|Q^{\Delta B=2}_{LL}(\mu_b)|B^0_s\rgl \equiv \frac 8 3 m^2_{B_s} 
f^2_{B_s}B(\mu_b)\; ,
\ee
where $B(\mu_b)$ is bag parameter of the $B_s$ meson and $f_{B_s}$ is its decay constant.

So far $B(\mu_b)$ has been computed by using lattice QCD. One of the major problems
with those computations is the following: the usual ligth quark lattice action breaks 
explicitely the chiral symmetry, 
which tremendeously complicates the renormalization procedure and matching to the 
continuum. To get around that problem we perform a first computation of $B(\mu_b)$ 
by using the lattice formulation of QCD in which 
the chiral symmetry is preserved at finite lattice spacing.\cite{overlap} On the other hand, 
it should be stressed that our heavy quark is static, as 
the currently available lattices do not allow to work with the propagating $b$ quark; 
so we chose to employ the heavy quark effective theory (HQET) in the static limit 
($m_b\ra \infty$). 

Four main steps implemented in our calculation are:
\newline
(1) Non perturbative computation of $\tilde{B} = 3/(8f^2_{B_s}) 
\lgl\overline{B^0_s}|\tilde
{Q}^{\Delta B=2}_{LL}|B^0_s\rgl$ in lattice HQET, where 
 $\tilde{Q}^{\Delta B=2}_{LL} = \bar{h}^i\gamma_\mu(1-\gamma^5) s^i 
 \bar{h}^j\gamma_\mu(1-\gamma^5)s^j$, $h$ being the static heavy quark field.
\newline
(2) Matching of $\lgl\overline{B^0_s}|\tilde{Q}^{\Delta B=2}_{LL}|B^0_s\rgl$ onto the 
continuum $\overline{{\rm MS}}$(NDR) scheme at 
NLO at the renormalization scale $\mu=1/a$, where $a$ is the lattice spacing. 
\cite{matchCONT}
\newline
(3) Running from $\mu=1/a$ to $m_b$ of $\lgl\overline{B^0_s}|\tilde{Q}^{\Delta B=2}(\mu)|B^0_s\rgl$ 
by the HQET anomalous dimension matrix, known to two loops accuracy in perturbation theory. 
\cite{run}$^,$\cite{truc}
\newline
(4) Matching of $\lgl\overline{B^0_s}|Q^{\Delta B=2}(\mu)|B^0_s\rgl$ onto their counterpart 
in QCD, in the $\overline{{\rm MS}}$(NDR) renormalization scheme at NLO. \cite{run} 

As a result of the above procedure we obtain: $\hat{B}^{\overline{MS},NLO}_{B_s}(m_b)=0.940(16)(22)$. \\
It can be seen from figure \ref{comp} that our value is larger than the previous static
result;\cite{static} this enhancement could be a proof that systematics are better 
controlled when subtractions are not needed. Our value is also somewhat larger than results 
obtained with the 
propagating heavy quark, which is likely due to the fact that we neglected $1/m_b$ 
corrections. Notice also that JLQCD collaboration showed that the errors due to quenching 
seem to be small.\cite{aoki1} \cite{aoki2} We also plan to address that issue by unquenching 
the $B^0_s-\overline{B^0_s}$ mixing amplitude 
in the static limit and by avoiding the subtraction procedure as well. The feasibility study 
by means of twisted mass QCD is underway.

\begin{center}

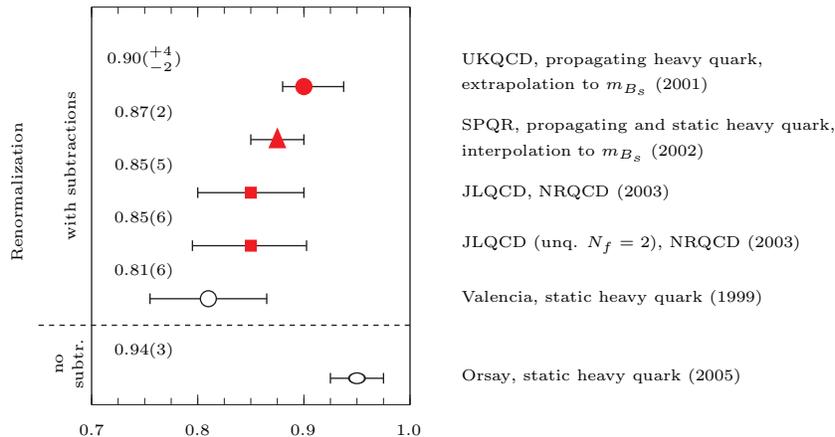
\begin{figure}[b]
\begin{picture}(40,-310)(-100,-320)

\LinAxis(0,-10)(120,-10)(3,4,-3.2,0,0.2)
\LinAxis(0,-160)(120,-160)(3,4,3.2,0,0.2)
\Line(0,-10)(0,-160) \Line(120,-10)(120,-160)
\DashLine(-20,-130)(120,-130){2}

\Text(0,-170)[]{\tiny{0.7}}
\Text(40,-170)[]{\tiny{0.8}}
\Text(80,-170)[]{\tiny{0.9}}
\Text(120,-170)[]{\tiny{1.0}}

\Text(20,-30)[]{\tiny{$0.90(^{+4}_{-2})$}}
\Text(20,-50)[]{\tiny{0.87(2)}}
\Text(20,-70)[]{\tiny{0.85(5)}}
\Text(20,-90)[]{\tiny{$0.85(6)$}}
\Text(20,-110)[]{\tiny{$0.81(6)$}}
\Text(20,-140)[]{\tiny{$0.94(3)$}}

\Line(90,-150)(110,-150)
\Line(90,-152)(90,-148)
\Line(110,-152)(110,-148)
\COval(100,-150)(2,3)(0){Black}{White}

\SetColor{Black}
\Line(22,-120)(66,-120)
\Line(22,-122)(22,-118)
\Line(66,-122)(66,-118)
\CCirc(44,-120){3}{Black}{White}

\Line(38,-100)(81,-100)
\Line(38,-102)(38,-98)
\Line(81,-102)(81,-98)
\CBoxc(60,-100)(4,4){Red}{Red}

\Line(40,-80)(80,-80)
\Line(40,-82)(40,-78)
\Line(80,-82)(80,-78)
\CBoxc(60,-80)(4,4){Red}{Red}

\Line(60,-60)(80,-60)
\Line(60,-62)(60,-58)
\Line(80,-62)(80,-58)
\CTri(67,-63)(70,-56)(73,-63){Red}{Red}


\SetColor{Black}
\Line(72,-40)(95,-40)
\Line(72,-42)(72,-38)
\Line(95,-42)(95,-38)
\CCirc(80,-40){3}{Red}{Red}

\Text(140,-150)[l]{\tiny{Orsay}, static heavy quark (2005)}

\Text(140,-120)[l]{\tiny{Valencia}, static heavy quark (1999)}

\Text(140,-100)[l]{\tiny{JLQCD (unq. $N_f=2$), NRQCD (2003)}}

\Text(140,-80)[l]{\tiny{JLQCD, NRQCD (2003)}}

\Text(140,-55)[l]{\tiny{SPQR, propagating and static heavy quark,}}
\Text(140,-65)[l]{\tiny{interpolation to $m_{B_s}$ (2002)}}

\Text(140,-30)[l]{\tiny{UKQCD, propagating heavy quark,}}
\Text(140,-40)[l]{\tiny{extrapolation to $m_{B_s}$ (2001)}}

\rText(-30,-80)[][l]{\tiny{Renormalization}}
\rText(-10,-70)[][l]{\tiny {with subtractions}}
\rText(-14,-145)[][l]{\tiny {no}}
\rText(-7,-145)[][l]{\tiny {subtr.}}

\end{picture}
\vspace{-5.3cm}
\caption{ \label{comp} Various lattice values of $\hat{B}^{\overline{MS}}_{B_s}(m_b)$;
unfilled symbols correspond to a computation made 
with a static heavy quark}
\end{figure}
\end{center}

\end{document}